# TRUSTED COMPUTING IN MOBILE ACTION

**Nicolai Kuntze, and Andreas U. Schmidt**

Fraunhofer-Insitute for Secure Information Technology SIT

Rheinstrasse 75

64295 Darmstadt, Germany

Phone: +49 6151 869 60227

Fax: +49 6151 869 224

{nicolai.kuntze,andreas.u.schmidt}@sit.fraunhofer.de

ABSTRACT

Due to the convergence of various mobile access technologies like UMTS, WLAN, and WiMax the need for a new supporting infrastructure arises. This infrastructure should be able to support more efficient ways to authenticate users and devices, potentially enabling novel services based on the security provided by the infrastructure. In this paper we exhibit some usage scenarios from the mobile domain integrating trusted computing, which show that trusted computing offers new paradigms for implementing trust and by this enables new technical applications and business scenarios. The scenarios show how the traditional boundaries between technical and authentication domains become permeable while a high security level is maintained.

# TRUSTED COMPUTING IN MOBILE ACTION

## 1 INTRODUCTION

Future business models have to support an evolving and diverse mobile ecosystem [1]. Due to the convergence of different mobile technologies like 3G, WLAN, Bluetooth, and WiMax the need for a new supporting infrastructure arises. This infrastructure must support usage scenarios which are independent of the underlying network and access model. In this heterogeneous and potentially de-centralised environment the underlying trust model is crucial for the success of old and novel business models. In a de-centralised trust model it is possible to relieve central components in an infrastructure from various tasks like trust establishment, accounting, or charging. This can help to reduce costs at the side of the infrastructure provider. By integrating certain functionalities into the mobile devices and establishing trust into this device it is possible to re-implement already known techniques, e.g. SIM-Lock, on a new security level and with enhanced possibilities.

Trusted Computing (TC) [2] offers new paradigms for implementing trust in computing platforms and between them by introducing a hardware-based trust anchor. This isolates the security functions from the other hardware and from the operating system and enables a clear design which is easy to evaluate with respect to security properties.

This concept paper discusses advanced scenarios for trusted computing use in the mobile domain. The common theme is the enabling of novel business scenarios [3, 4] centred on the mobile network operator (MNO). It should be noted that our scenarios are using trusted computing in a different way than in Digital Rights Management (DRM), which is often considered as the sole use for TC. The presented scenarios are binding the economic value to a particular instantiation of the physical trust anchor. If the trust anchor breaks, only a limited damage can occur as the damage is restricted in space and time. On the converse, if a single trust anchor breaks in the DRM scenario, the protected digital good can be converted into an unprotected version which can afterwards be freely distributed on a large scale, causing heavy monetary losses to its owner.

In Section 2 TC is introduced comprehensibly as the underlying technology. Section 3 presents functional restriction as the first major application and an anonymous, prepaid mobile phone as a second application derived from functional restriction. Section 4 examines machine-to-machine (M2M) communication from the viewpoint of mobile and de-centralised business models. From this second major theme we derive a point of acceptance scenario and a method to mediate physical access to buildings or facilities.

## 2 TRUSTED COMPUTING

Trusted computing has the goal to establish trust in computing platforms. The Trusted Computing Group (TCG) has defined a family of standards to accomplish this task. A so called Trusted Platform Module (TPM) is used as a trust anchor offering the ability to securely store and use asymmetric keys. Moreover, a TPM in cooperation with an appropriate operating environment can issue to a third party assertions about the trustworthiness of the computing platform, e.g., signal that the platform is in a secure state. Such a TPM enhanced system is called trusted platform.

The TPM offers so-called shielded capabilities which are protecting internal data structures by controlling their use. In the presented applications we will use two particular features of a trusted platform. First, the creation, storage, and usage of keys and their protection by the TPM, and second the ability to create a trust measurement which can be used to assert a certain state towards a third party. The TPM is designed for using asymmetric keys and is equipped with a physical random number generator, and key generation component which creates RSA key pairs. The key generator is designed as a protected capability and the created private keys are kept in a shielded capability. The key generation component can also be used for creation of symmetric keys but the TPM can only offer storage and protection for the latter.

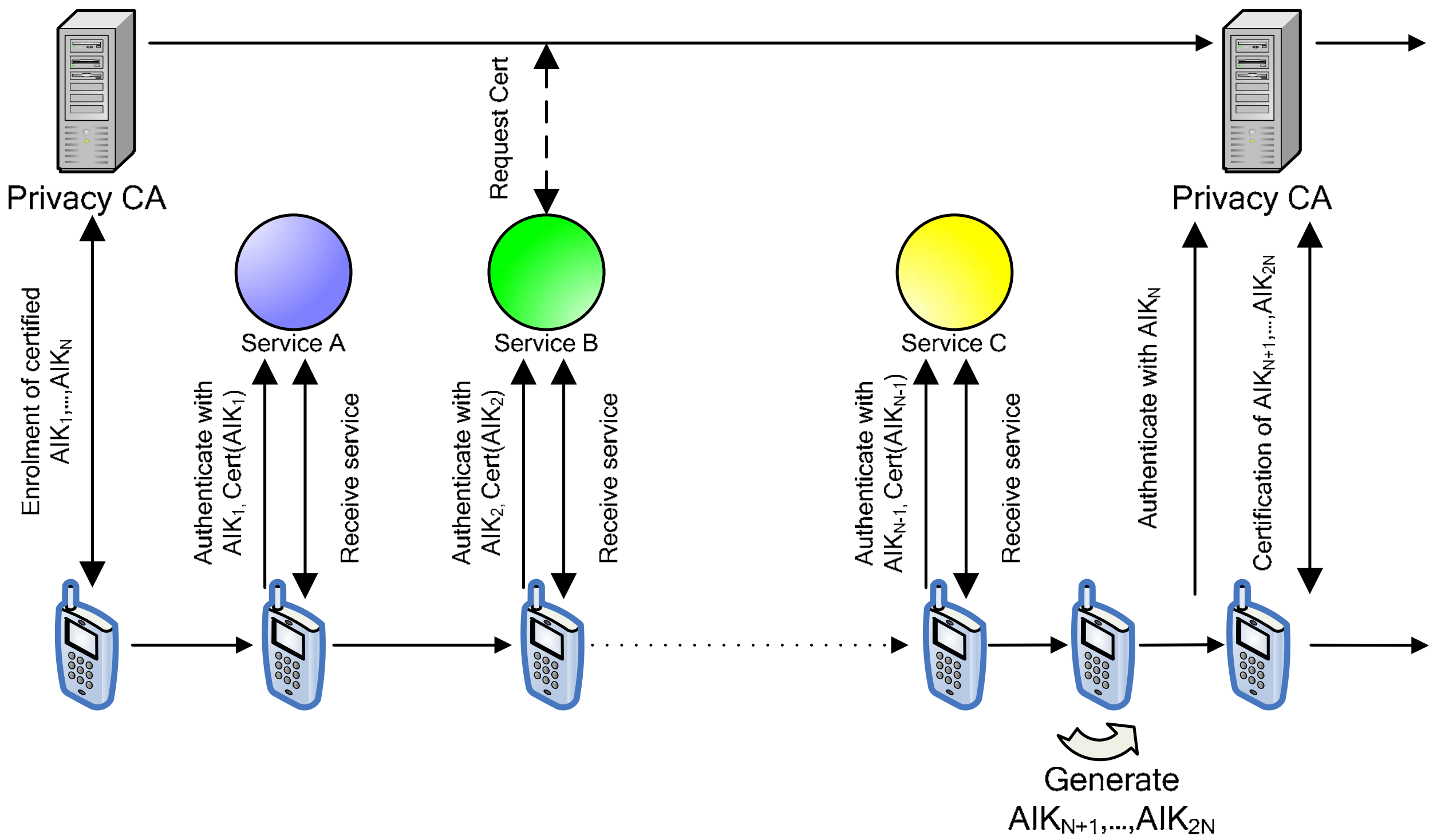

*Figure 1: Use of AIKs for one-time authentication credentials.*

In the context of trust measurement the TPM offers Platform Configuration Registers (PCR) which are implemented as 160-bit storage locations. The PCRs are hardware base for attesting the system state to a third party. Performing such an attestation means to signal to the remote party that the device (1) is unchanged and the platform (2) is in a trustworthy configuration. Proving the integrity requires to record and measure the system state beginning at the system boot. For this, the concept of a trusted boot process is introduced. As the anchor of this process an initial component starts and measures itself. This initial integrity value is saved in a log and reported to the TPM which stores it in a PCR. After this, the initial component loads and measures the BIOS. The resulting integrity value is also reported and logged before the BIOS starts. This process of measurement, reporting, and logging is iterated by each component. The trusted OS finally performs these tasks for every software component loaded. All integrity values reported to the TPM are added to the PCR by applying the SHA-1 hash algorithm.

During the so-called remote attestation process the verifier receives the log and a signed PCR value as well as the certificates to verify the signature. PCRs are signed by special signing keys whose operation is limited to signing PCRs. These Attestation Identity Keys (AIKs) are created by the TPM. The AIKs have to be linked to the TPM which is uniquely identified by the Endorsement Key (EK). Signing the AIK with the unique EK would raise many privacy concerns as it would be possible to create attestation profiles for a device. The TCG has therefore introduced a Trusted Third Party (or privacy CA) which is able to state that a certain AIK belongs to a TPM and that this TPM and the platform on which it resides are conforming to certain constraints. During the attestation the verifier is presented a certificate from this privacy CA and has to decide if this privacy CA is trustworthy. Based on this certificate the verifier can test requirement (1) from above. The requirement (2) is tested by comparing the signed PCR value with the provided log. An attacker can alter the log but not the PCR value. Hence it should be possible to recalculate the PCR value with the data provided by the log. The verifier has to verify the logged integrity values with reference values which may make it in general complicated to perform this process.

In the mobile domain there are certain constraints different from the PC domain. The devices are rather restricted in computing power, energy supply, and memory. Due to these special requirements a special Mobile Phone Working Group of the TCG designs a mobile TPM (mTPM). To establish usage of TC in the praxis this mTPM has to offer a device sided verifier which can offer assertions for the integrity values. Without this ability each mobile device has to contact a central service provider for checking the integrity values.

## 2.1 Privacy, Identity, Pseudonymity, …

Issues of privacy are central to the applicability and adoption of TC technology and have been main points of critique on the work of the TCG. The AIK and its certificate prove assertion (1) from above. There is no relation to a dedicated user as the AIK is meant to communicate the trustworthiness of a single instantiation of a trusted platform. If it can be assumed that there is a one to one relation between device and owner, as it is common in the mobile domain (e.g. SIM card), the AIK can be used identifying a user.

Figure 1 elucidates a rather simple approach to protect user's privacy in service access scenarios. It uses the TC concept of a privacy CA supporting pseudonymity in remote attestation in a slightly odd way. Here, a TC privacy CA takes a double role, not only certifying AIKs for remote attestation but also aiding in pseudonymous authentication for service access. The privacy CA (PCA) thus acquires features of an identity provider. One may think of a collaboration of service providers who ask the PCA to issue a dedicated certificate for their authentication decisions. When a device carries out remote attestation with one of these service providers using this certificate, it concurrently authenticates itself and qualifies for service access.

A key feature of this concept is that accumulation of customer data or profiles is impeded by the use of one AIK per transaction. The AIK/certificate pair is analogous to a one-time PIN/TAN combination. When all but one AIKs are used up, the TPM of the device generates a new list of AIKs and obtains certificates for them from the PCA using the last remaining one of the old list.

The remote attestation process offers pseudonymity for the users in the attestation process. As this is not always enough TCG has established Direct Anonymous Attestation (DAA) [5, 6] which authenticates the AIK by a protocol related to bit commitment and zero knowledge proofs [7]. Although DAA is not practically relevant today due to high requirements with respect to computational power and lack of implementations, its features make it a promising technology on the medium term. DAA is joint work of IBM's Zurich research Labs, Intel, and HP, and is basically a lightweight version of Zurich Research Lab's idemix system [8].

DAA enables a trusted platform to proof to a remote party the presence of a valid TPM and the result of trust measurements without revealing the platform's identity. During roll-out a TPM is assigned to a certain trust domain but during operation no central authority is needed. The DAA method uses a different AIK per transaction and authenticates the AIK with the DAA protocol proper (basically replacing certificate-based authentication by DAA in Figure 1).

It should be noted that using AIKs for authentication service toward services is actually an abuse, since they are originally only meant to sign the PCR values. But as we have shown they are rather handy for the former secondary use. The talk [9] discusses how AIK-based authentication can be efficiently implemented. There is a lot of ongoing research internationally, and in particular in the EU in the OpenTC project [10, 11]. One particular aim of OpenTC is to provide an open platform for service access based on TC.

# 3 FUNCTIONAL RESTRICTION

In the relationship between network operator and customer in the mobile domain, the standard form of customer retention exerted by the MNO is SIM-lock, a crude form of functional restriction of mobile devices bonding them to a certain MNO. Using TC, a finer grained functional discrimination of mobile devices becomes possible. Depending on business models, various client functions of the device can be restricted to certain, more or less privileged customer groups. Device management, of which functional discrimination is an important instance, is viewed by the industry as a fundamental application area of TC [12].

For an MNO, it is cost-efficient to produce a single product line with many appearances to the end-user, rather than marketing a multitude of makes and models as customary today. Second, the up- and downgrading of functionalities can be implemented dynamically, without physical access to the device. To the user, the relative seamlessness with which device control operates is an ergonomic benefit and allows for better customisation and even personalisation. The means to implement functional restrictions is provided by the trusted boot process and operating system of the trusted platform. TC assures that the device belongs to a certain, restricted group, and that it is in a state where only the allowed functions are enabled. The set of functions to be managed can be pre-configured and the dynamic control effected via simple changes of parameters. The enforcement level of this approach is stronger as compared to SIM-lock precisely because the trusted platform's base operation software is tamper resistant. Based on this assurance, the MNO can deliver specific services or content only to the restricted group privy to it. Thus functional restriction provides the foundation on the client side for further service discrimination, policy enforcement, and DRM proper.

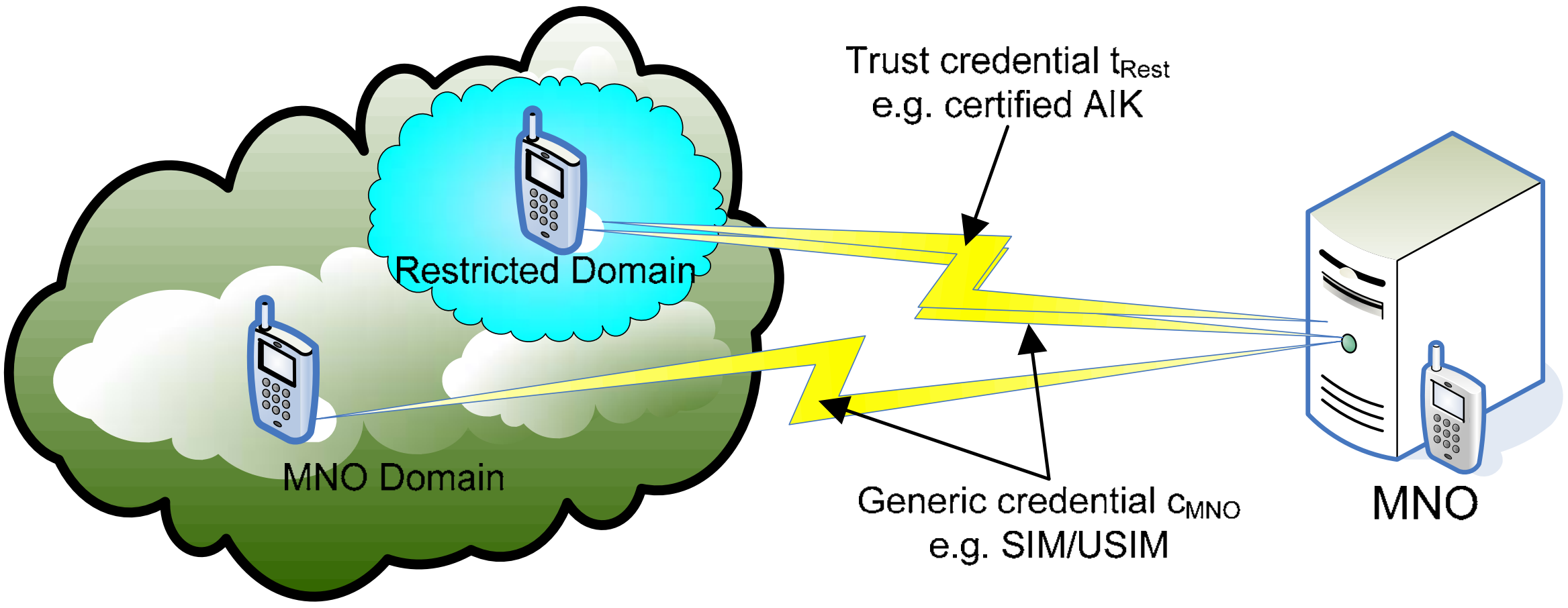

*Figure2: Restriction to a sub-domain by trust credentials.*

From an abstract point of view [13], functional restriction is based on a more general restriction process which appears in one way or another on each of the scenarios in the present paper. As shown in Figure 2, restriction means creating a subgroup of agents *viz.* devices privy to more services or restricted in functionality. Restriction is a basic authentication function by which an agent is assigned to a sub-domain, e.g., within the domain of an MNO. It works by a combination of credentials or authentication tokens. All devices in the larger domain still use a so-called *generic credential* $c_{MNO}$, usually bearing the individual identity of the device, to gain basic access to the communication network; assignment to the sub-domain is then done by transmitting an additional credential. This *trust credential* $t_{Rest}$ can for instance be realised as signed AIK(s), certified by a privacy CA which may or may not be identical to the MNO. In this case, a remote attestation process is carried out between device and MNO through the MNO's own network. Alternatively, trusted software on the device, i.e., software which is included in the trusted boot process and of which the fact that it is present and not adulterated is shown via remote attestation, can provide own, special credentials, e.g., cryptographic secrets of any kind. From the trustworthy assertions that a TPM can deliver, restriction needs 1. Presence of a live and unaltered TPM. This can for instance be carried out using a challenge-response method using the TPM's endorsement credential. And 2. The integrity of the system and its components. This property is ascertained through trust measurements and communicated via remote attestation.

The additional security and in effect higher trust in agents of the restricted group enables those clients to access additional content or services. This is in fact the classical scenario used to enforce copyright protection through digital rights management (DRM). It may be more the rule than the exception that $t_{Rest}$ provides stronger authentication than $c_{MNO}$. From a security viewpoint, it would then be the preferable to use only $t_{Rest}$ to authenticate restricted agents. However, this is often not practical, e.g., when the communication channel through which $t_{Rest}$ is conveyed is only available after authentication by $c_{MNO}$.

The possibility for the MNO (maybe in co-operation with the privacy CA) to check the consistency of $c_{MNO}$ with $t_{Rest}$ makes the subgroup more resilient against cloning attacks on the generic credential. This kind of attack is not uncommon in the mobile sector, see [14]. A proper choice of enrolment method and time for the trust credential is essential for the validity of the additional trust provided by the restriction operation. If both credentials are impressed on the agents independently of each other, i.e., not both under the control of the MNO, then, e.g., resilience against cloning attacks is restricted. Since he cannot associate the two credentials belonging to an individual agent, he can at best avoid granting two agents with identical $c_{MNO}$ service access by using a first-come-first-served approach. Higher cloning-resilience can be achieved if a single authority individualises both credentials and controls their deployment to the devices.

Privacy in functional restriction is limited by concept, since communication and in particular authentication is under the control of the MNO. Measures for privacy enhancement in service access based on restriction can nevertheless protect certain transaction data from the MNO. That is the MNO notices only that a transaction was carried out from a certain device with a particular third party service provider, but not the particular details of it (such as good and amount purchased).

A combination of functional restriction by TC with mobility suggests itself. This enables for instance location-based scenarios with high security requirements. Trusted client software on a mobile device can enable and disable certain features if it based on position information derived from the current cell of the mobile network or GPS: This can be used as a counter measure against (industrial) espionage in sensitive areas. Conversely, sensitive personal data, e.g., pictures, can be transferred from a central service to the device only when it enters a pre-defined (home-) cell.

## 3.1 Prepaid Mobile Phone

A prepaid mobile phone can be implemented utilising functional restriction. The phone carries in its trusted storage area a running total which is decremented by a trusted software. While the initial access to the mobile network is still established using SIM authentication, a TC-based remote attestation and authentication then yield assurance to the MNO that the running total is nonzero, upon which access to the network's communication services can be granted. This releases the MNO from operating (or paying for) a centralised accounting. In this scenario the user acquiring the device can remain anonymous. This is not legal in the EU, where every SIM is registered with an individual but may be a viable option in other countries. The technical implementation scenario for a TC-based prepaid phone we envisage as a 'Gedankenexperiment' is rather fancy since it annuls the usual SIM authentication completely (it would even obviate the transmission of SS7 authentication triplets).

Assume that the mobile device is equipped with two trusted software clients. A virtualised SIM (VSIM) for basic network access and a prepaid client (ppC) for service access and control of the running total. The presence and integrity of both components is measured and during the trusted boot process and recorded in the associated log. VSIM possesses a group credential ppIMSI that replaces the ordinary International Mobile Subscriber Identity (IMSI) with which a mobile phone usually logs on to an MNO's network. ppIMSI may be a set of actual, reserved IMSIs from which the device choose one at random for log on, to mitigate (unlikely) concurrency conflicts. VSIM thus provides a common identity for the pre-assigned group of prepaid phones which appear identical to the MNO. Authentication by VSIM is therefore of course weaker than SIM authentication.

As normal, subsequent addressing from network to device is then carried out by assigning a dedicated Temporary Mobile Subscriber Identity (TMSI) to the device. At this point the device is still restricted to a quarantine sub-network not allowing any communication with another party but the MNO. The device then carries out a remote attestation using an AIK and transmitting in particular the trust measurements of VSIM and ppC. This completes the modified network log on depicted in Step 1 in Figure 3, after which the MNO can allow general service access for this device. The outlined procedure is similar in philosophy to the designs of the Trusted Network Connect Working Group of the TCG [15].

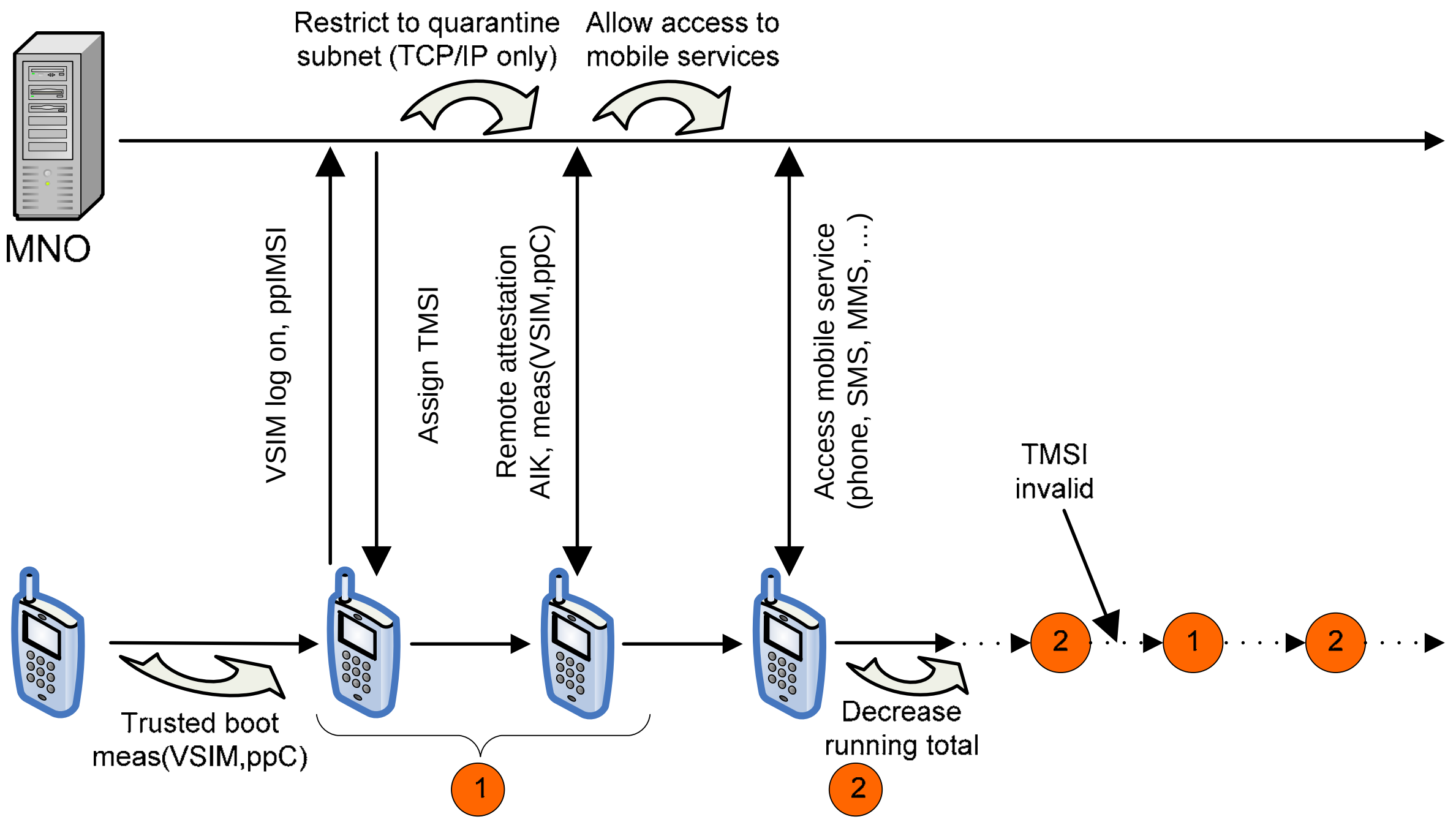

*Figure 3: Anonymous prepaid access to mobile communication*

In Step 2 the device requests mobile services, receives them, and decreases its running total accordingly. Depending on security requirements, the MNO may demand a remote attestation for every single service access, which can then be done with one-time AIKs as explained in Section 2.1. When the TMSI becomes invalid – as, e.g., an MNO might want to assign to TMSIs for prepaids a limited lifetime for security reasons – Step 1 has to be repeated.

If the devices are cheap enough they may actually be designed as one-way. Recharging once the running total is depleted can be done in various ways, e.g., employing a third party charging service provider, similar to the scenario described below in Section 4.

## 4 MACHINE-TO-MACHINE COMMUNICATION

A trusted mobile device can be used as a relay for the transmission of sensitive data in a machine-to-machine (M2M) scenario. Thus the MNO can provide a service to a third party, e.g., for advanced monitoring systems for control and maintenance of technical installations such as district heating and cooling systems, water supply and wastewater, or remote metering of such installations. Consumption of electricity, water and heating can be remotely read in each apartment of a block and automatically transferred to the administrative system. Similarly in-house lighting, burglary alarms, camera surveillance, booking of a laundry room, control of pre-heaters, supervision of windows, doors, stove, and so on can be controlled. As is clear from this extensive list of applications, trusted M2M communication can be a base infrastructure for general facility management.

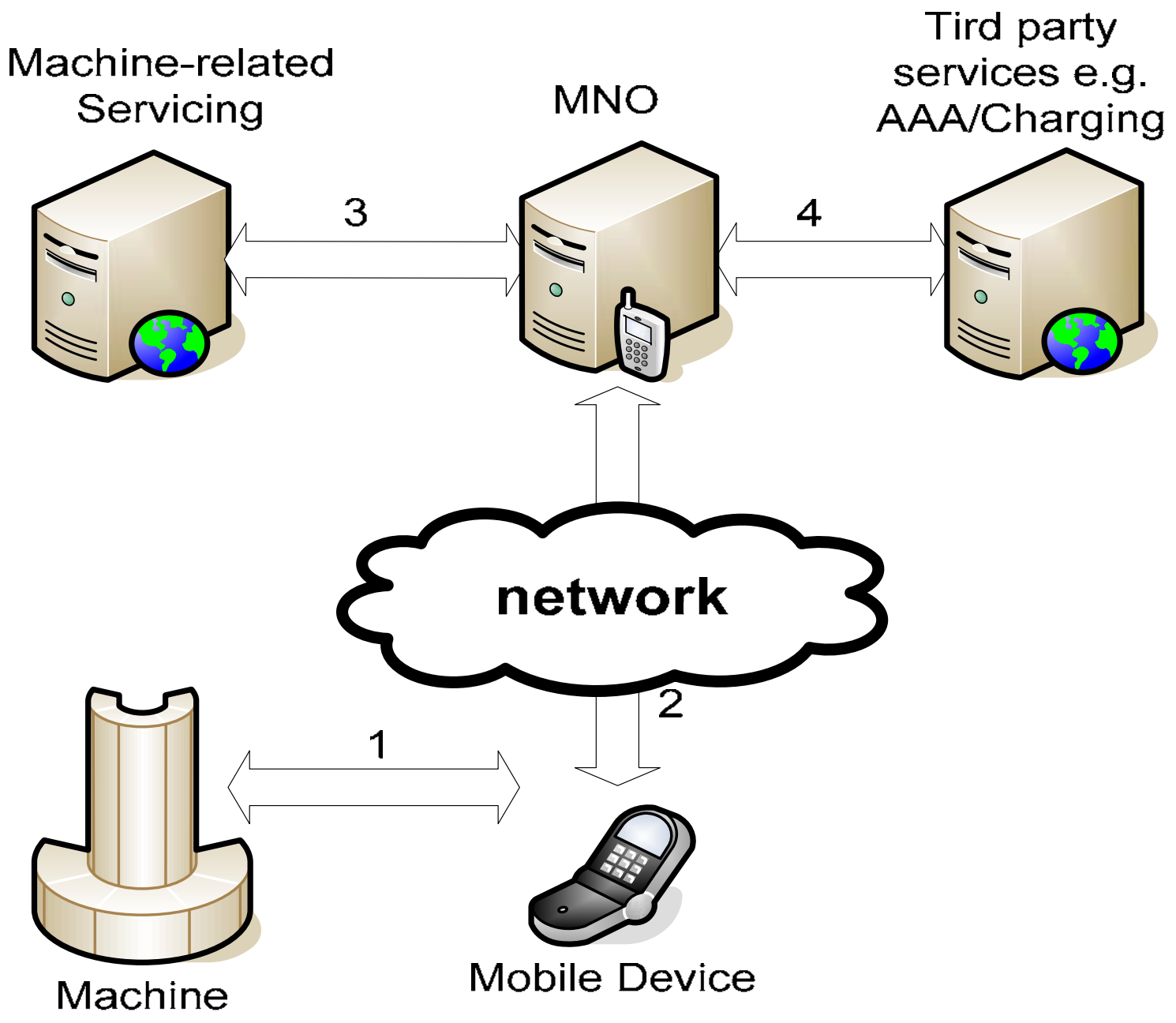

*Figure 4: Generic trusted M2M scenario.*

As seen in Figure 4 the machine uses the mobile device as a toehold in the communication with the MNO. The mobile device enables another device to utilise its connection to the MNO for authentication, authorisation, and accounting (AAA). The mobile device can either ask on behalf of the machine or just relay AAA requests and related data. From the viewpoint of cost-effectiveness it is very advantageous that no GSM or UMTS module is required in the machine. The communication can be performed using a near range communication module like Bluetooth or IRDA. TC can provide mutual authentication and even data protection in this scenario.

Second, and most importantly, the authentication of the mobile device is performed by the well-known SIM-card authentication. The authentication between the mobile device and the machine is based on secrets offered by the mobile device's TPM. This is used to authenticate and ultimately authorise the customer at the charging unit. The mobile network operator offers in such a scenario services both to his own clientele as well as to the machine proprietor, and can thus profit from his established infrastructure and large subscriber base. By separating the duties [16, 17] between a charging unit and the MNO, additionally certain privacy options arise could enable minimal need to know policies. For instance, the MNO need not know the identity of the single machine involved.

A special instance of M2M communication, in which the MNO takes part in accounting and charging processes of a third party, is a point of sales (POS) scenario. The benefits for the vendor that arise in this scenario basically stem from the trust relationship that is mediated between MNO and POS by the mobile device. In this way the MNO provides payment services as well as authorisation control for the vendor. In extended service scenarios, the customer's mobile devices can as well be utilised to transfer valuable information to the POS, e.g., updated price and commodity lists, or firmware. As a side effect the statistics over the delivered goods could be forwarded to the vendor so that the logistics can be optimised.

A coarse description of the purchase process at the operational level is as follows. A user with a TPM-equipped mobile device wants to purchase a soft drink from a likewise trust-enabled POS. While the user still makes up her mind on her taste preferences, device and POS initiate a trusted communication session using remote attestation and transport layer encryption. Device and POS thus achieve mutual assurance that they are in an unaltered, trustworthy state, and begin to exchange price lists and payment modalities (1 in Figure 4).

After the user selects a good and confirms his choice at his device, signed price and payment processing information is transferred to the MNO (2 in Figure 4). After verifying the signatures and optionally informing the good’s vendor and a payment service provider (3 and 4 in Figure 4), the MNO sends a signed acknowledgement to the mobile device, which relays it to the POS, where it is verified and the good is delivered.

The individual identities of POS and device (and therefore its owner) need not be revealed in the purchase process. Authentication is carried out and trust is established by referral. That is the respective providers of identities of the two domains in question – the MNO and the POS vendor – vouch for the identities of their agents.

Authentication of the POS can be achieved by at least two ways protecting the anonymity of the POS. First, the mobile device connects directly the POS owner and verifies a credential. This requires an initial trust of the mobile device into the POS owner. Alternatively the mobile device asks the MNO about the identity of the POS. The MNO can verify the POS credential in some way and acknowledge the request. The identity of the POS is revealed to the MNO by this operation.

Using a PCA on side of the POS owner, the POS can change its identity after a certain time or use some identities in parallel providing at least pseudonymity as detailed in Section 2.1. This prevents compilation of purchase data by the MNO. Therefore each POS can acquire as many AIKs as necessary thus protect its own identity (duration of validity of the certificates can be chosen as small as possible to avoid the necessity of revocation lists). A similar implementation would by possible using DAA without necessitating the online requests to a privacy CA.

### 4.1 Data protection in M2M scenarios

In the economic environment it is often required to implement a certain kind of minimal need to know principle to protect each communication partner from disclosing corporate secrets or informational assets. For example the charging service provider may not be meant to know the list of goods, the stock, price list, or other details of a certain POS owner. Furthermore, the MNO should not know how many points of sale are owned by a certain company, where these POS' are located, and generally how the POS owner’s business is doing. With respect to these data protection targets and independently of implementation details and variants, the POS scenario can be analysed as follows. During a purchase operation

(i) the mobile device authenticates itself at the POS.

(ii) To verify the authentication token offered by the mobile device the POS connects to the POS owner infrastructure where the decision of acceptation is made. This decision is made based on a trust relationship between POS owner and an authentication provider (e.g. a charging provider). Alternatively the POS requests this directly at the authentication provider. In both cases the POS owner gets no knowledge about the identity of the customer.

(iii) After the authentication the purchase process is performed at the POS. The resulting data containing the billing information like authentication token, good identifier, and price, are transferred to the POS owner, where a special data package is generated for the charging provider. This package needs in principle only to contain the authentication token and the grand total to charge. Hence the charging provider does not gain any information about the good, quantities, and qualities sold by the POS owner, nor essentials of his infrastructure.

(iv) After the confirmation of the charging the POS owner acknowledges the purchase and the POS vending machine delivers the good.

Steps (iii) and (iv) can be de-centralised in an alternative approach, and therefore as well be performed by the POS. The POS in this case requests the confirmation from the charging provider and afterwards requests the acknowledgement by the POS owner.

From the viewpoint of the charging provider relations between costumers and POS owning companies can arise, e.g., in the form of special offers and rebates. These marketing strategies can remain closed to the other involved parties. In particular there is no possibility to connect individual consumers to the purchased products even for the charging provider who cannot, e.g., build a customer profile using his data alone.

From the point of view of the network operator all traffic is encrypted so it is not possible for her to distinguish between a POS and any other device. Privacy concerns could arise if the network operator and authentication provider are integrated in one party. In this scenario it could be possible to compile user profiles and reconstruct buying patterns.

In the mutual authentication of the mobile device and the POS, end-to-end security is achieved. In particular, remote attestation of the device toward the POS yields to the latter the assurance that full price and lists of available goods will not leave the device. This is possible since the device is proved to be in a trustworthy state, belongs to the MNO's domain, and, optionally, end-to-end encryption between POS and device is established (e.g., using TC functionality).

### 4.2 Physical Access Control and Facility Management

As the most advanced scenario we consider the control of physical access to buildings using standard, TC-enabled mobile devices. In fact, this can be extended to a full-fledged facility management utilising the MNOs infrastructure, yielding large potential savings to the building's management. This scenario is very attractive since it enables a much more efficient deployment process than traditional facility management and access control. For instance specialised hardware tokens are no longer needed and out-of-the-box solutions can be envisaged. The scenario neatly combines functional restriction with M2M functionalities supported by TC.

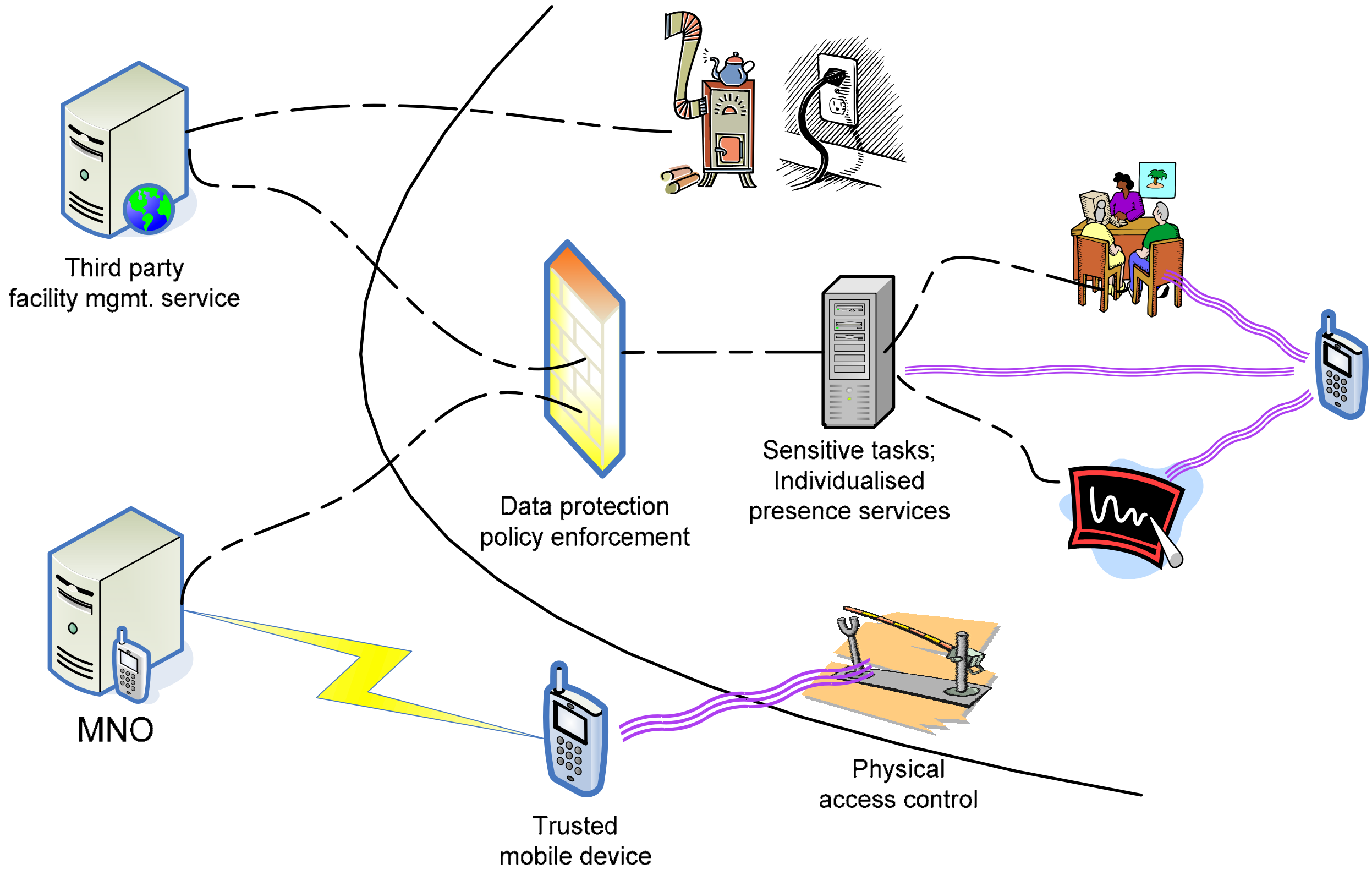

*Figure 5: Distributed access control and facility management utilising trusted mobile devices*

The scenario is depicted in Figure 5. It should be clear from the following high level description that there are many implementation variants of it. Assume a company moves in to an existing building, e.g., on lease, that already has management of base infrastructures like power supply, heating, and air-conditioning in place. These services are thought of as being out-sourced to a third party service provider. The company, of course, needs more functionality, like presence services, room reservation, and physical access control, many of which use sensitive data. In implementing such services, a main goal is to minimise infrastructure costs. This can be supported by utilising TC-enabled mobile devices, which can be bought and supplied by the company itself, subsidised by an MNO, or even be the private devices of the employees. As a main feature of the devices, some kind of short-range communication is used to de-centralise communication within the facility like in the generic M2M scenario above.

A primary task is distributed physical access control. In security-sensitive areas this is usually realised using dedicated smart tokens as replacements of keys. Access gates and doors are commonly connected to the internal IP network to enable, e.g., access logs. Using TC-equipped mobile devices instead, the devices authenticate themselves to the gate, e.g., as in POS scenario. This can include an updating of the access policy or online verification of the access rights via the MNO network. During access control the employees device undergoes functional restriction, for example cameras are disabled and sending of MMS is suppressed (or at least controlled, e.g., in collaboration with the MNO). We assume that the MNO is connected to some central server owned by the company which controls security related and sensitive tasks. All connections to this server should go through a policy enforcer which prevents sending sensitive data to the outside.

Within the building, the device helps in more refined services as mentioned above. It uses short range communication to communicate with white-boards, internal room control, coffee dispensers, and so on. All these terminals can do without network capabilities as in the POS scenario, since communication to the internal server is done via the devices (authentication included). For certain tasks, like preventing shutting down power in a room when a midnight meeting takes place, the internal server can communicate with the external facility management provider, in which case a data protection policy must again be applied.

# 5 CONCLUSION

We have shown that trusted computing has a wide area of application in the mobile domain. From an architectural viewpoint the presented scenarios lever the classical client/server relationship between mobile device and its user on the one hand and the network and the services provided through it on the other hand. Scenarios become possible in which services are exchanged between these two and other business parties. It is an intriguing question how this will affect business models in the mobile domain, in particular if it actually enables new ones.

Applications for trust-enabled mobile networks and devices exist at all levels of trust from the provisioning of basic functions like network and service access to transactions demanding highest security such as establishment of ad hoc contracts and Service Level Agreements, and possibly even (legally binding, qualified) electronic signatures. We have shown that personal data can be protected in a reasonable way by TC technology in these applications.

For the MNO, the prospect to federate the identities of millions of subscribers of mobile networks with other providers of goods and services is rather attractive. De-centralised authentication through the trusted agents has the benefit of enhanced resilience and availability of service access. It can also be a base for de-centralised authorisation and ultimately de-centralised business models, such as super-distribution of virtual goods from agent to agent. Building de-centralised trust structures like Webs of trust [18] is another notable option.